# Proximity Full-Text Search by Means of Additional Indexes with Multi-Component Keys: in Pursuit of Optimal Performance

Alexander B. Veretennikov[1(✉)][0000-0002-3399-1889]

[1] Ural Federal University, Yekaterinburg, Russia
`alexander@veretennikov.ru`

**Abstract.** Full-text search engines are important tools for information retrieval. In a proximity full-text search, a document is relevant if it contains query terms near each other, especially if the query terms are frequently occurring words. For each word in a text, we use additional indexes to store information about nearby words that are at distances from the given word of less than or equal to the *MaxDistance* parameter. We showed that additional indexes with three-component keys can be used to improve the average query execution time by up to 94.7 times if the queries consist of high-frequency occurring words. In this paper, we present a new search algorithm with even more performance gains. We consider several strategies for selecting multi-component key indexes for a specific query and compare these strategies with the optimal strategy. We also present the results of search experiments, which show that three-component key indexes enable much faster searches in comparison with two-component key indexes.

## 1 Introduction

A search query consists of several words. The search result is a list of documents containing these words. In [1], we discussed a methodology for high-performance proximity full-text searches and a search algorithm. With the application of additional indexes [1], we improved the average query processing time by a factor of 94.7 when queries consist of high-frequency occurring words.

In this paper, we present the following new results.

We present a new search algorithm in which we can improve the performance even more than it was improved in [1].

We present the results of search experiments that prove that three-component key indexes can be used to improve the average query execution time by up to 15.6 times in comparison with two-component key indexes when queries consist of high-frequency occurring words.

In modern full-text search approaches, it is important for a document to contain search query words near each other in order to be relevant to the context of the query, especially if the query contains frequently occurring words. The impact of the term-proximity is integrated into modern information retrieval models [2-5].

Words appear in texts at different frequencies. The typical word frequency distribution is described by Zipf's law [6]. An example of words' occurrence distribution is shown in Fig. 1. The horizontal axis represents different words in decreasing order of their occurrence in texts. On the vertical axis, we plot the number of occurrences of each word.

The full-text search task can be solved with inverted indexes [7-9]. With ordinary inverted indexes, for each word in the indexed document, we store in the index the record ($ID$, $P$), where $ID$ is the identifier of the document and $P$ is the position of the word in the document. Let $P$ be an ordinal number of the word in the document.

For proximity full-text searches, we need to store the ($ID$, $P$) record for all occurrences of any word in the indexed document. These ($ID$, $P$) records are called "postings".

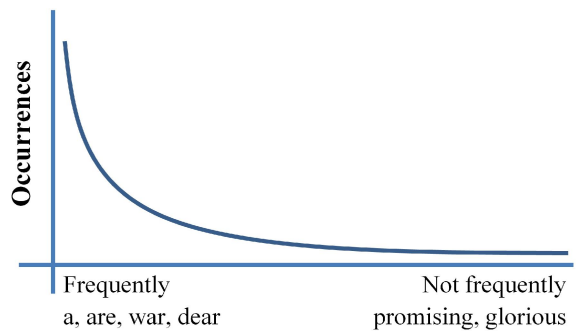

**Fig. 1.** Example of a word frequency distribution.

Therefore, the query search time is proportional to the number of occurrences of the queried words in the indexed documents. Consequently, to evaluate a search query that contains high-frequency occurring words, a search system needs much more time (see Fig. 1, on the left side) than a query that contains ordinary words (see Fig. 1, on the right side).

A full-text query is a "simple inquiry", and accordingly [10], to prevent the interruption of the thought continuity of the user, the query results must be produced within two seconds. In this context, we present the following problem. It is common to have a full-text search engine that can usually evaluate a query within 1 sec. of time. However, it works very slowly, for example, requiring 10-30 sec., for a query that contains frequently occurring words.

We can illustrate this problem by the following example. We downloaded pgdvd042010.iso from the Project Gutenberg web page, which contains their files as of April 2010, and we indexed its content using Apache Lucene 7.4.0 and Apache Tika 1.18. We indexed approximately 64 thousand documents with a total length of approximately 13 milliard characters (a relatively small number). We indexed all words. Then, we evaluated the following queries using the equipment from section 4.1 of the current paper:

"Prince Hamlet"~4 – this search took 172 milliseconds, and

"to be or not to be"~4 – this search took 21 seconds.

The suffix "~4" instructs Lucene to search such texts in which the queried words contain no more than 4 other words between them.

To improve the search performance, early-termination approaches can be applied [11-12]. However, early-termination methods are not effective in the case of proximity full-text searches [1]. It is difficult to combine the early-termination approach with the integration of term-proximity information into relevance models.

Another approach is to create additional indexes. In [13-14], the authors introduced some additional indexes to improve the search performance, but they only improved phrase searches.

With our additional indexes, an arbitrary query can be evaluated very quickly.

In this paper, we present a new and more effective approach that extends the method from [15]. In the new approach, we try to select the optimal configuration of multi-component key indexes for a specific query. The major extension is shown in the "3.3 Index selection" section, and the results of new experiments are presented.

## 2    Lemmatization and Lemma Type

### 2.1    Word Type

In [16], we defined three types of words.

**Stop words**: Examples include "and", "at", "or", "not", "yes", "who", "to", "war", "time", "man" and "be". In a stop-word approach, these words are excluded from consideration, but we do not do so. In our approach, we include information about all words in the indexes.

We cannot exclude a word from the search because a high-frequency occurring word can have a specific meaning in the context of a specific query [1, 14]; therefore, excluding some words from consideration can induce search quality degradation or unpredictable effects [14].

Let us consider the query example "who are you who". The Who are an English rock band, and "Who are You" is one of their songs. Therefore, the word "Who" has a specific meaning in the context of this query.

**Frequently used words**: These words are frequently encountered but convey meaning. These words always need to be included in the index. Examples include "beautiful", "red", and "hair".

**Ordinary words**: This category contains all other words. Examples include "glorious" and "promising".

### 2.2 Lemmatization

We employ a morphological analyzer for lemmatization. For each word in the dictionary, the analyzer provides a list of numbers of lemmas (i.e., basic or canonical forms). For a word that does not exist in the dictionary, its lemma is the same as the word itself. Some words have several lemmas. For example, the word "mine" has two lemmas, namely, "mine" and "my".

We use a combined Russian/English dictionary with approximately 200 thousand Russian lemmas and 92 thousand English lemmas.

We define three types of lemmas: stop lemmas, frequently used lemmas and ordinary lemmas. We sort all lemmas in decreasing order of their occurrence frequency in the texts. We call this sorted list the *FL*-list. The number of a lemma in the *FL*-list is called its *FL*-number. Let the *FL*-number of a lemma $w$ be denoted by $FL(w)$.

The first *SWCount* most frequently occurring lemmas are stop lemmas. The second *FUCount* most frequently occurring lemmas are frequently used lemmas. All other lemmas are ordinary lemmas. *SWCount* and *FUCount* are the parameters. We use *SWCount* = 700 and *FUCount* = 2100 in the experiments presented.

If an ordinary lemma, $q$, occurs in the text so rarely that $FL(q)$ is irrelevant, then we can say that $FL(q) = \sim$. We denote by "$\sim$" some large number.

### 2.3 Index Type

We create indexes of different types for different types of lemmas. Let *MaxDistance* be a parameter that can take a value of 5, 7 or even greater.

The expanded ($f$, $s$, $t$) index or three-component key index [1, 17] is the list of occurrences of the lemma $f$ for which lemmas $s$ and $t$ both occur in the text at distances that are less than or equal to the *MaxDistance* from $f$.

We create an expanded ($f$, $s$, $t$) index only for the case in which $f \leq s \leq t$. Here, $f$, $s$, and $t$ are all stop lemmas. Each posting includes the distance between $f$ and $s$ in the text and the distance between $f$ and $t$ in the text.

The expanded (*w*, *v*) index or two-component key index [18-20] is the list of occurrences of the lemma *w* for which lemma *v* occurs in the text at a distance that is less than or equal to the *MaxDistance* from *w*.

The lemma types considered are as follows: for *w*, frequently used, and for *v*, frequently used or ordinary. Each posting includes the distance between *w* and *v* in the text.

Other types of additional indexes are described in [1].

## 3 A New Search Algorithm

### 3.1 The Search Algorithm General Structure

Our search algorithm is described in Fig. 2. Let us consider the search query "who are you who". After lemmatization, we have the following query:

[who] [are, be] [you] [who]. The word "are" has two lemmas in our dictionary.

With *FL*-numbers: [who: 293] [are: 268, be: 21] [you: 47] [who: 293].

To use three-component key indexes, this query must be divided into two subqueries [1]:

*Q1*: [who: 293] [are: 268] [you: 47] [who: 293], and

*Q2*: [who: 293] [be: 21] [you: 47] [who: 293].

We can say that lemma "who" > "you" because *FL*(who) = 293, *FL*(you) = 47, and 293 > 47. We use the *FL*-numbers to establish the order of the lemmas in the set of all lemmas.

In [1], we defined several query types depending on the types of lemmas that they contain and the different search algorithms for these query types. The query does not need to be divided into a set of subqueries for all query types.

In this paper, we consider subqueries that consist only of stop lemmas.

After step 2, we evaluate the subqueries in the loop.

After all subqueries are evaluated, their results need to be combined into the final result set.

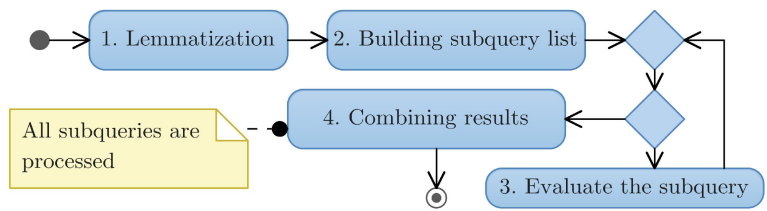

**Fig. 2.** UML diagram of the search algorithm general structure.

### 3.2 Subquery Evaluation

The algorithm for the subquery evaluation, when the subquery consists of only stop lemmas, is described in Fig. 3.

We need to select the three-component key indexes required to evaluate the subquery. For all selected indexes, we need to create an iterator object. The iterator object for the key (*f*, *s*, *t*) is used to read the posting list of the (*f*, *s*, *t*) key from the start to the end. The iterator object, *IT*, has the method *IT.Next*, which reads the next record from the posting list.

The iterator object, *IT*, has the property *IT.Value*, which contains the current record (*ID, P, D1, D2*). Consequently, *IT.Value.ID* is the *ID* of the document containing the key, and *IT.Value.P* is the position of the key in the document.

For the two postings of *A* = (*A.ID*, *A.P, A.D1, A.D2*) and *B* = (*B.ID*, *B.P, B.D1, B.D2*), we define that *A* < *B* when one of the following conditions is met: *A.ID* < *B.ID* or (*A.ID* = *B.ID* and *A.P* < *B.P*). The records (*ID, P, D1, D2*) are stored in the posting list for the given key in increasing order.

The goal of the *Equalize* procedure is to ensure that all iterators have an equal value of *Value.ID* = *DID*. Afterwards, we can perform the search in the document with identifier *Value.ID*. The *Equalize* procedure is described in [1].

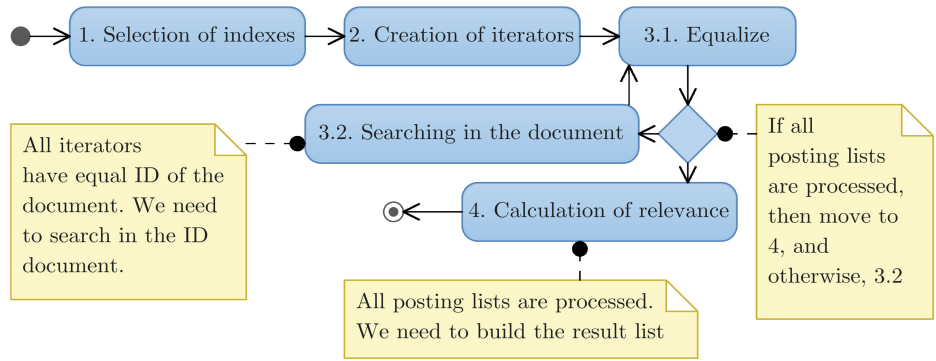

**Fig. 3.** UML diagram of a subquery evaluation.

### 3.3 Index Selection

To evaluate the subquery, we need to select keys for the three-component key indexes. The key selection can be performed in different ways for different performance outcomes. We propose now four different approaches. The simple and effective principles, which are defined as the second and third approaches below, can significantly increase performance in comparison with the original first approach.

**The First Approach**
The first approach is proposed in [15]. The query can be divided into a set of three-component keys. Let the first three lemmas of the query define the first key. Let the next three lemmas of the query define the second key, and so on.

For the cases when the length of the query is not an exact multiple of 3, the last key is always defined by the last three lemmas of the query.

All selected keys must be normalized.

For example, let us consider the subquery [who] [are] [you] [who]. We can use the keys (who, are, you) and (are*, you*, who).

For any three stop lemmas, *f*, *s* and *t*, we have the (*f*, *s*, *t*) index only for the case in which $f \leq s \leq t$. We call the (*f*, *s*, *t*) key with the aforementioned condition the normalized key. The normalized keys here are (you, are, who) and (you*, are*, who).

Let us consider the search query "Who are you and why did you say what you did" and its subquery [who] [are] [you] [and] [why] [do] [you] [say] [what] [you] [do].

In fact, we can find this query in Cecil Forester Scott's novel "Lord Hornblower".

We can use the (who, are, you), (and, why, do), (you, say, what), and (what*, you, do) indexes. The normalized keys are (you, are, who), (and, do, why), (you, what, say), and (you, what*, do). We mark "what" by "*" in the last key to denote that this lemma has already been taken into account by a previous key.

**The Second Approach**

The idea of the second approach is the following. Let query *Q* be a list of lemmas. In any case, we need to use the most frequently occurring lemma for a component of a three-component key. This lemma will be the first component of the first key. However, we can minimize the number of postings to read by selecting the least frequently occurring lemmas, which we can find in the query, as the other two components of the key. After we form the first key, we can apply the aforementioned logic to select the following key using the remaining lemmas of the query, etc.

When we form a key, we always need to select the lemmas at different indexes in *Q*. For this, we will "mark" an item of the *Q* as "used" when we select it.

We perform the following steps in the loop.

1. If all elements of *Q* are "used", then we break the loop.
2. We select a lemma *f* with index *x* in the *Q* with the following conditions:
   a. *x* is not used,
   b. lemma *f* is the most frequently occurring lemma that satisfies the previous condition.
3. We mark *x* as "used".
4. We try to select a lemma *s* with index *y* in the query with the following conditions:
   a. *y* is not used,
   b. *s* is the least frequently occurring lemma that satisfies the previous condition.
5. If we cannot select a lemma in the previous step, then we select lemma *s* with index *y* in the query with the following conditions (additionally, *s* is marked with * in the key):
   a. *y* is not equal to *x*.
   b. *s* is the least frequently occurring lemma that satisfies the previous condition.
6. We mark "y" as "used".
7. We try to select a lemma *t* with index *z* in the query with the following conditions:
   a. *z* is not used,
   b. *t* is the least frequently occurring lemma that satisfies the previous condition.

8. If we cannot select a lemma in the previous step, then we select lemma *t* with index *z* in the query with the following conditions (additionally, *t* is marked with * in the key):
   a. *x* is not equal to *z*, and *y* is not equal to *z*.
   b. *t* is the least frequently occurring lemma that satisfies the previous condition.
9. We mark *z* as "used".
10. We create a three-component key (*f*, *s*, *t*) and include it in the list of keys.

We present two examples of the second approach.

Let us consider the subquery *SQ1* = [who] [are] [you] [who].

With *FL*-numbers, we have the following: [who: 293] [are: 268] [you: 47] [who: 293].

We select "you" as the first component of the key because it is not "used" and has the most occurrence frequency in the texts, that is, the lowest *FL*-number of 47.

Then, we select "who" as the second component of the key and "who" as the third component of the key. We have the key (you, who, who) and the normalized key (you, who, who). The indexes 0, 2 and 3 are used.

Then, we select the remaining "are" as the first component of the second key. All of the indexes are "used" now. Thus, we select "who" and the second "who" as the second and the third components, respectively, and we have the (are, who*, who*) key.

Let us consider the subquery *SQ2* of another query = [who] [are] [you] [and] [why] [do] [you] [say] [what] [you] [do].

With *FL*-numbers we have the following: [who: 293] [are: 268] [you: 47] [and: 28] [why: 528] [do: 154] [you: 47] [say: 165] [what: 132] [you: 47] [do: 154].

We select "and: 28" as the first component of the first key and "why: 528" and "who: 293" as the second and the third components, respectively. Then we select "you: 47", "are: 268", and "say: 165" for the second key. Then, we select "you: 47", "do: 154", and "do: 154" for the third key. Then we select "you: 47", "what: 132", and "why*: 528" for the last key. The normalized keys are (and, who, why), (you, say, are), (you, do, do), and (you, what, why*).

It is important to remember that we need to divide the query into parts if we have an index with a small *MaxDistance* value. Any part of the divided query must have a length that is less than or equal to the *MaxDistance*. To evaluate the subquery *SQ2* without division, we need at least a *MaxDistance* = 11.

If we consider the discussion about relevance from [21], then the lengths of the parts must be less than the *MaxDistance* to some extent. For example, if the *MaxDistance* = 5, then we can limit the length of each part by the number 4, and we can consider the following division: **[**[who] [are] [you] [and]**]**, **[**[why] [do] [you] [say]**]**, **[**[what] [you] [do]**]**. Each of the parts must be independently evaluated, and after that, the results of these evaluations must be combined.

**The Third Approach**

We have an important observation regarding the second approach. When we select a high frequently occurring lemma as the first component of the key and some less frequently occurring lemma as the other component of the key, this can significantly reduce the number of the postings to read. In the second approach, we select both the second and the third components of the key as the least frequently occurring lemmas. However, what if the query contains several high frequently occurring lemmas and a small number or even only two relatively low frequently occurring lemmas? In this case, it may be useful to not "spend" all of the least frequently occurring lemmas for the one key but to distribute them somehow between several keys.

In the first step, we determine the number of required keys. Let us have a subquery of length *n*. We need $k = n / 3$ indexes, which number we need to round up. The query is a list of lemmas; thus, each item of the list has its index in the list. For any component of each key, we need to select an index of a lemma in the query. When we select an index for a specific key, we "mark" the index as "used"; thus, it cannot be used for another key.

For each key, we perform the following. We select the most frequently occurring unmarked lemma in the subquery as the first component of the key, and the least frequently occurring unmarked lemma in the subquery as the third component of the key. We perform this for all of the keys. Then, we must select the second component for every key.

For each key we perform the following. If we have "unmarked" indexes, then we select the least frequently occurring unmarked lemma in the subquery as the second component of the key; otherwise, we select the least frequently occurring lemma in the subquery, whose index is not used by any component of the current key (in the latter case, the component of the key is marked with *).

Let us consider *SQ2* again. We need to define four keys.

In the first step, we define the first and the third components for each key.

We select "and: 28" as the first component of the first key and "why: 528" as the third component of the first key. We select "you: 47" and "who: 293" for the second key and "you: 47" and "are: 268" for the third key. We select "you: 47" and "say: 165" for the last key.

Then, we select "do: 154" as the second component for the first key. We select "do: 154" for the second key, "what: 132" for the third key and "why*: 528" for the last key. The normalized keys are (and, do, why), (you, do, who), (you, what, are), and (you, say, why*).

**The Fourth Approach and Analysis**

In the fourth approach, we consider all possible variants of key selection. In this approach, we need the ability, which we have, to estimate the count of postings for any three-component key. In this case, if we consider all possible variants of the key selection, we can select an optimal variant of the key selection (with the least number of postings to read that is required for the query evaluation).

The problem of this approach is as follows. With an increase in the length of the query, the number of variants for the key selection increases very quickly. However, this approach can be used for short queries and for analysis of optimality of other approaches.

In [15], we presented results for the first approach. In this new paper we present results for the other approaches, which are more promising.

We postpone for now the question of how to work with duplicates among the lemmas of the query.

The presented approaches can be applied not only for three-component key indexes, but also for *n*-component key indexes, *n* > 3, if they are needed. They can also be used in a reduced way for 2-component indexes.

### 3.4   Search in the Document

The algorithm of searching in the document is described in Fig. 4.

Let *DID* be an argument of the "Search in the document" procedure. Let us define that *DID* is the identifier of the current document.

The main difference between the search algorithm from [1] and the new search algorithm is described here.

For any lemma in the search query, we create an intermediate list of postings in memory. For example, let us consider the three-component index (you, are, who) and its iterator object. We create three intermediate posting lists: *IL(*you), *IL*(are), and *IL*(who). To fill these intermediate posting lists, we need to read postings from the (you, are, who) iterator object.

A record from the (you, are, who) iterator object has the format (*ID*, *P*, *D1*, *D2*), where *ID* is the identifier of the document, *P* is the position of "you" in the document, *D1* is the distance from "are" to "you" in the text, and *D2* is the distance from "who" to "you" in the text.

If the lemma "are" occurs in the text after the lemma "you", then D1 > 0; otherwise, D1 < 0.

If the lemma "who" occurs in the text after the lemma "you", then D2 > 0; otherwise, D2 < 0.

We need to read from the iterator object all records with *ID* = *DID*.

We can produce three records from the (*ID*, *P*, *D1*, *D2*) record.

We need to store the (*P*) record in the *IL*(you) intermediate posting list.

We need to store the (*P* + *D1*) record in the *IL*(are) intermediate posting list.

We need to store the *(P* + *D2)* record in the *IL*(who) intermediate posting list.

Let us consider the key (you, what*, do) with a lemma marked by "*". In this case, we create only two intermediate posting lists, namely, *IL*(you) and *IL*(do). The (what*) component is already taken into account by a previous key.

For each lemma of the subquery, since we have the intermediate posting list, the search is straightforward and similar to the search in the ordinary inverted file.

Additionally, an intermediate posting list is a kind of iterator object. The intermediate posting list object, *IL,* has the method *IL.Next*, which reads the next record from the posting list.

The intermediate posting list object, *IL,* has the property *IL.Value,* which contains the current record (*P*), where *P* is the position of its lemma in the document.

Let *IL.Value* be equal to *SIZE_MAX* when all records are read from the *IL* object, where *SIZE_MAX* is some large number.

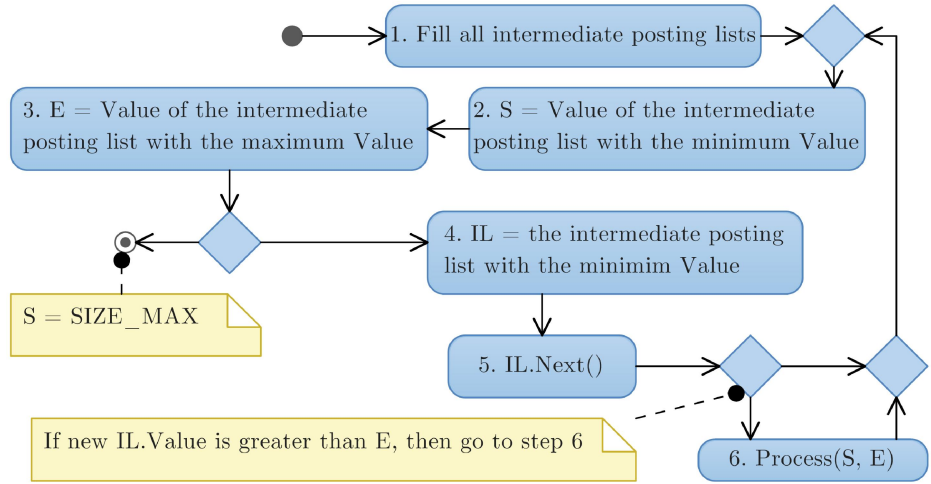

**Fig. 4.** UML diagram of searching in a document.

In the loop, we perform the following steps.

1. Let *MinIL* be the intermediate posting list with a minimum value of *Value*.
2. Let *S* = *MinIL.Value*.
3. Let *MaxIL* be the intermediate posting list with a maximum value of *Value*.
4. Let *E* = *MaxIL.Value*.
5. If there are no more records in *MinIL*, then exit from the search.
6. Execute *MinIL.Next*().
7. If *MinIL.Value* > *E*, then execute *Process*(*S*, *E*).
8. Go to step 1.

The *Process*(*S*, *E*) procedure adds the (*DID, S, E*) record into the result set. *S* is the position of the start of the fragment of text within the document that contains the query. *E* is the position of the end of the fragment of text within the document that contains the query.

### 3.5 Intermediate Posting List Data Ordering

The records (*P*) must be stored in an intermediate posting list for the given lemma in increasing order. For this requirement, the following problem arises.

Consider the text "to be or not to be or". Let the position of a word in the text be its ordinal number starting with zero. When we create the three-component key index, the following records must be stored for the key (to, be, or). The records in the format (position of "to", position of "be", position of "or") are presented below.

(to, be, or): (0, 1, 2), (0, 5, 6), (4, 1, 2), and (4, 5, 6).

From this posting list, we can create the following three intermediate posting lists.

(to): 0, 0, 4, 4; (be): 1, 5, 1, 5; and (or): 2, 6, 2, 6.

Only for the first component of the key is the intermediate posting list ordered in increasing order.

Please note that the postings in the three-component key index will actually be encoded in the (*ID*, *P*, *D1*, *D2*) format. For the (to, be, or) key, we will write the following posting list: (to, be, or): (*ID*, 0, 1, 2), (*ID*, 0, 5, 6), (*ID*, 4, –3, –2), and (*ID*, 4, 1, 2).

To solve the aforementioned problem, we create two binary heaps [22]. We create the first binary heap for the second component of the key. We create the second binary heap for the third component of the key.

Therefore, we will create the (be) binary heap and the (or) binary heap.

We limit the binary heap length by *MaxDistance* × 2.

When we need to read postings from the (to, be, or) posting list, we perform the following in a loop.

1. Read the next posting (*ID*, *P*, *D1*, *D2*) from the posting list (to, be, or).
2. Write (*P*) into the (to) intermediate posting list.
3. Write (*P* + *D1*) into the (be) binary heap.
4. Let *M* be the first (the minimum element) of the (be) binary heap. If the length of the (be) binary heap is greater than *MaxDistance* × 2 or if the distance between *M* and the new element (*P* + *D1*) is greater than *MaxDistance* × 2, then remove the first element from this binary heap, and write it into the (be) intermediate posting list.
5. Write (*P* + *D2*) into the (or) binary heap.
6. Let *M* be the first (the minimum element) of the (or) binary heap. If the length of the (or) binary heap is greater than *MaxDistance* × 2 or if the distance between *M* and the new element (*P* + *D2*) is greater than *MaxDistance* × 2, then remove the first element from this binary heap, and write it into the (or) intermediate posting list.
7. Go to step 1.

Let us consider a key (*f*, *s*, *t*) and its posting list, *L*. We create three intermediate posting lists and two binary heaps to proceed as follows.

1. Intermediate posting list *F* for *f*.
2. Intermediate posting list *S* and binary heap *SH* for *s*.
3. Intermediate posting list *T* and binary heap *TH* for *t*.

Let us introduce the methods *PopMin, Min* and *Length* of a binary heap object. The *PopMin* method returns the minimum element from the binary heap and removes this element from the binary heap. The *Length* method returns the length of the binary heap. The *Min* method returns the minimum element from the binary heap but does not change the binary heap.

In Fig. 5, we present the UML diagram of the posting list *L* reading process.

After all postings from *L* are read, we need to write all elements from the binary heaps to their intermediate posting lists.

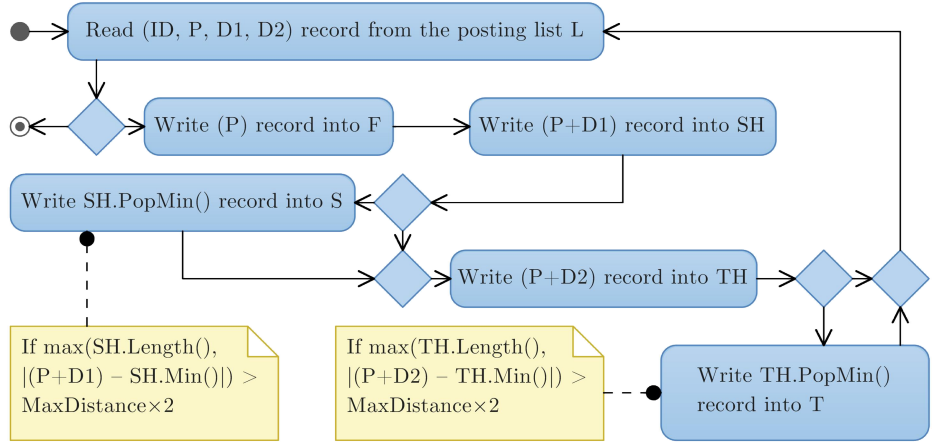

**Fig. 5.** UML diagram of the posting list reading process.

### 3.6 Advantages of the New Algorithm

The new algorithm may require a smaller amount of posting lists to evaluate a search query in comparison with the algorithm from [1] and, therefore, provides faster searches. It also allows for a more flexible key selection.

### 3.7 Computational Complexity

Let $Q$ be a subquery of $m$ lemmas. Let $n$ be the total number of postings to read for the query evaluation.

For each posting, we need to use it in the *Equalize* procedure. In [1], the author states that the cost of such usage is $O(\log(m))$.

For each posting, we need to add it to the three intermediate posting lists. The cost of this process is $O(\log(MaxDistance))$ (see section 3.5).

For each posting, we need to use it when searching in a document. The cost of this process is $O(\log(m))$ (see section 3.4).

The final cost of the subquery evaluation is

$O(n \cdot (\log(m) + \log(MaxDistance)) = O(n \cdot \log(\max(m, MaxDistance)))$.

## 4 Search Experiments

### 4.1 Search Experiment Environment

All search experiments were conducted using a collection of texts from [1]. The total size of the text collection was 71.5 GB. The text collection consisted of 195 000 documents of plain text, fiction and magazine articles. We used *MaxDistance* = 5, *SWCount* = 700, and *FUCount* = 2100. The search experiments were conducted using the experimental methodology from [1].

We assume that in typical texts, words are distributed similarly, in accordance with Zipf's law [6]. Therefore, the results obtained with our text collection will be relevant to other collections.

We used the following computational resources:

CPU: Intel(R) Core(TM) i7 CPU 920 @ 2.67 GHz.

HDD: 7200 RPM. RAM: 24 GB.

OS: Microsoft Windows 2008 R2 Enterprise.

We created the following indexes.

Idx1: the ordinary inverted index without any improvements, such as NSW records [1, 19]. The total size was 95 GB.

Idx2: our indexes, including the ordinary inverted index with the NSW records and the ($w$, $v$) and ($f$, $s$, $t$) indexes, where *MaxDistance* = 5. The total size was 746 GB.

Please note that the total size of each type of index includes the size of the repository (indexed texts in compressed form), which was 47.2 GB.

### 4.2 Search Results

There are 975 queries, and all queries consisted only of stop lemmas. The query set was selected as in [1]. All searches were performed in a single program thread. We searched all queries from the query set with different types of indexes to estimate the performance gains of our indexes. The query length was from 3 to 5 words.

Studies by Jansen et al. [23] have shown that queries with lengths greater than 5 are very rare. In [23], query logs of a search system were analyzed, and it was established that queries with a length of 6 represent approximately 1% of all queries and that fewer than 4% of all queries had more than 6 terms.

We performed the following experiments.

SE1: all queries are evaluated using the standard inverted index Idx1.

SE2.1: all queries are evaluated using Idx2 and the algorithm from [1].

SE2.2: all queries are evaluated using Idx2, the novel algorithm presented in this paper and that in [15] with the key selection based on the first approach.

SE2.3: all queries are evaluated using Idx2 and the novel algorithm presented in this paper with the key selection based on the second approach.

SE2.4: all queries are evaluated using Idx2 and the novel algorithm presented in this paper with the key selection based on the third approach.

SE2.5: all queries are evaluated using Idx2 and the novel algorithm presented in this paper with the key selection based on the fourth approach.

Average query times:

SE1: 31.27 sec., SE2.1: 0.33 sec., SE2.2: 0.29 sec., SE2.3: 0.24 sec., SE2.4: 0.24 sec., and SE2.5: 0.27 sec.

Average data read sizes per query:

SE1: 745 MB, SE2.1: 8.45 MB, SE2.2: 6.82 MB, SE2.3: 6.2 MB, SE2.4: 6.16 MB, and SE2.5: 5.79 MB.

Average numbers of postings per query:

SE1: 193 million, SE2.1: 765 thousand, SE2.2: 559 thousand, SE2.3: 423 thousand, SE2.4: 419 thousand, and SE2.5: 411 thousand.

We improved the query processing time by a factor of 94.7 with the SE2.1 algorithm, by a factor of 107.8 with the SE2.2 algorithm, and by a factor of 130 with the SE2.3 and SE2.4 algorithms (see Fig. 6).

In SE2.5 we observed a slight increase in the average query execution time, because this time includes checking all the possible combinations of the three-component key selection. We used the SE2.5 results only for analysis of the effectiveness of SE2.3, and SE2.4 in relation to SE2.2 (see later); therefore, SE2.5 is excluded from Fig. 6.

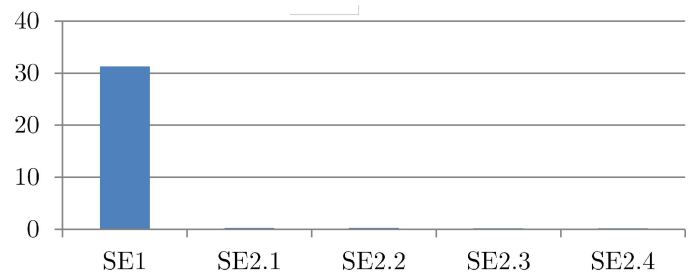

**Fig. 6.** Average query execution times for SE1, SE2.1, SE2.2, SE2.3, and SE2.4 (in seconds).

The left-hand bar shows the average query execution time with the standard inverted indexes. The subsequent bars show the average query execution times with our indexes using the SE2.1, SE2.2, SE2.3 and SE2.4 algorithms. Our bars are much smaller than the left-hand bar because our searches are very quick.

We improved the data read size per query by a factor of 88 with SE2.1, by a factor of 109.2 with SE2.2 and by a factor of 120 with SE2.3 and SE2.4 (see Fig. 7).

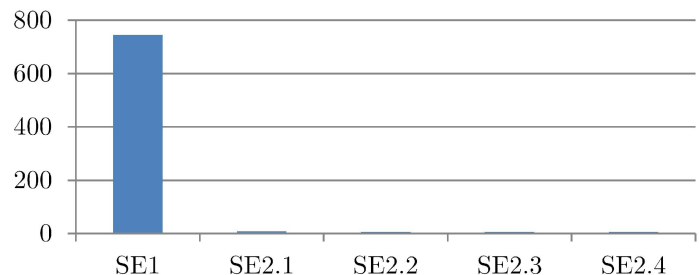

**Fig. 7.** Average data read sizes per query for SE1, SE2.1, SE2.2, SE2.3 and SE2.4 (MB).

The left-hand bar shows the average data read size per query for SE1. The subsequent bars show the average data read size per query for SE2.1, SE2.2, SE2.3 and SE2.4.

We show how SE2.3 and SE2.4 outperform SE2.2 in Fig. 8.

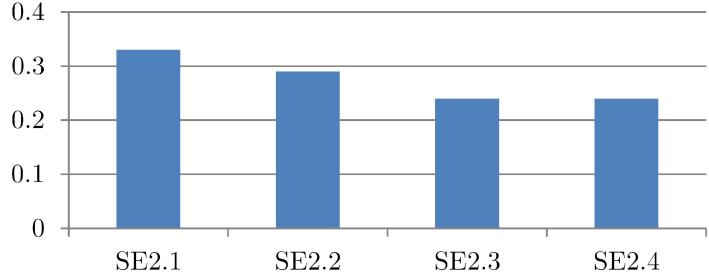

**Fig. 8.** Average query execution times for SE2.1, SE2.2, SE2.3 and SE2.4 (in seconds).

We show the average number of postings to read per query for SE2.1, SE2.2, SE2.3, SE2.4 and SE2.5 in Fig. 9. We observe that SE2.3 and SE2.4 have similar effectiveness in comparison with SE2.5, which is the optimal key selection. We also observe how SE2.3 and SE2.4 outperform the original SE2.2 method.

SE2.3 and SE.2.4 have equal performance on average; however, we have examples of queries that have significantly different execution times for the SE2.3 and SE2.4 approaches. If we have information about the posting list length for every key, then it will be good to quickly check both SE2.3 and SE2.4 strategies before evaluating a specific query.

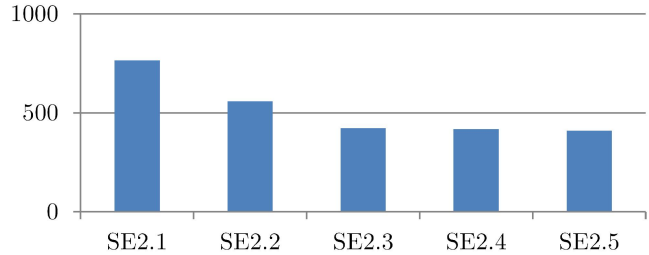

**Fig. 9.** Average numbers of postings to read per query for SE2.1, SE2.2, SE2.3, SE2.4 and SE2.5 (in thousands).

### 4.3  Comparison Between Three-Component Key Indexes and Two-Component Key Indexes

We created another additional index especially for this experiment.

Idx3: two-component key indexes (*w*, *v*), where *MaxDistance* = 5, *SWCount* = 0, and *FUCount* = 700. The total index size is 275 GB.

In this case, for any two lemmas, *w* and *v*, where $w \leq v$, $FL(w) < 700$, and $FL(v) < 700$, we have a two-component key index (*w, v*).

Each posting in this index includes the distance between *w* and *v* in the text.

Such *w* and *v* lemmas are stop lemmas for Idx2.

We performed the following experiment:

SE3: all 975 aforementioned queries were evaluated using Idx3, and the new algorithm presented in this paper is adapted for two-component key indexes.

In SE3, we processed the same query set that we already processed in SE2.1, SE2.2, SE2.3, and SE2.4, but we used two-component key indexes instead of three-component key indexes.

Average query times: SE3: 3.75 sec. (see Fig. 10).

Average data read sizes per query: SE3: 105.17 MB.

Average number of postings per query: SE3: 12 million 761 thousand.

In this experiment, we compared SE2.1, SE2.2, SE2.3 and SE2.4 against SE3. We improved the query processing time by a factor of 11.36 with the SE2.1 algorithm, by a factor of 12.93 with the SE2.2 algorithm, and by a factor of 15.6 with SE2.3 and SE2.4 in comparison with the two-component key index (SE3) case (see Fig. 10).

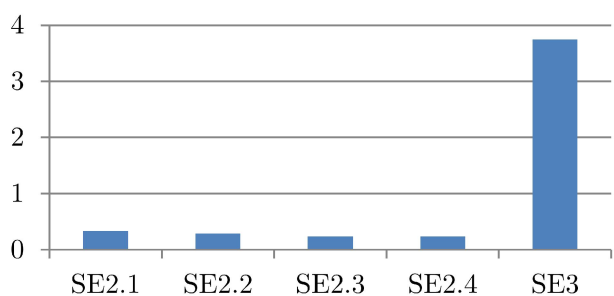

**Fig. 10.** Average query execution times for SE2.1, SE2.2, SE2.3, SE2.4 and SE3 (in seconds).

The left-hand bar shows the average query execution time with the three-component key indexes using the algorithm from [1]. The three center bars show the average query execution time with the three-component key indexes using the new algorithm described in this paper. The right-hand bar shows the average query execution time with the two-component key indexes.

The bars that related to the three-component key indexes are much smaller than the right-hand bar because the three-component key indexes enable much quicker searches than the two-component key indexes.

This experiment shows that three-component key indexes **by an order of magnitude are more effective** than the two-component indexes when the queries that consist of stop lemmas are evaluated.

We improved the data read size per query by a factor of 12.44 with SE2.1, by a factor of 15.42 with SE2.2 and by a factor of 16.96 with SE2.3 and SE2.4 in comparison with the two-component key index (SE3) case (see Fig. 11).

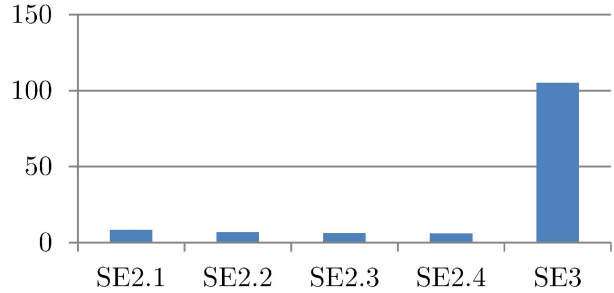

**Fig. 11.** Average data read sizes per query for SE2.1, SE2.2, SE2.3, SE2.4 and SE3 (MB).

The left-hand bar shows the average data read size per query with SE2.1. The subsequent bars show the average data read size per query with SE2.2, SE2.3, SE2.4 and SE3.

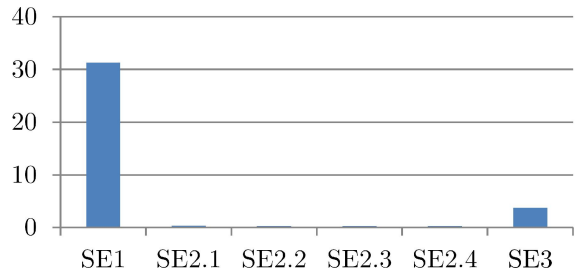

**Fig. 12.** Average query execution times for SE1, SE2.1, SE2.2, SE2.3, SE2.4 and SE3 (in seconds).

We show the average query execution time for all experiments in Fig. 12.

The left-hand bar shows the average query execution time with the standard inverted indexes. The four subsequent bars show the average query execution times with the three-component key indexes for the SE2.1, SE2.2, SE2.3 and SE2.4 algorithms. The right-hand bar shows the average query execution time with the two-component key indexes in the SE3 experiment.

## 5 Conclusion and Future Work

A query that contains high-frequency occurring words induces performance problems. To solve these performance problems and to satisfy the fastidious demands of the users, we developed and elaborated three-component key indexes.

In this paper, we investigated searches with queries that contain only stop lemmas. Other query types were studied in [18-19, 21]. As we discussed in [1], three-component key indexes are an important and integral part of our comprehensive full-text search methodology, which comprises three-component key index search methods and other search methods from [18-19, 21].

In this paper, we have introduced an optimized algorithm for full-text searches in comparison with [1]. These algorithms are novel, and no alternative implementations exist. We have analyzed different strategies for multi-component key selection for a specific query in pursuit of the best and optimal strategy.

We have presented the results of experiments showing that when queries contain only stop lemmas, the average time of the query execution with our indexes is 130 times less (with the *MaxDistance* = 5) than that required when using ordinary inverted indexes.

We have presented the results of experiments showing that when queries contain only stop lemmas, the average time of the query execution with our indexes is 15.6 times less (with the *MaxDistance* = 5) than that required when using two-component key indexes.

Using the last experiment, we diligently prove that three-component indexes are stupendous and cannot be replaced by two-component key indexes. This is the reason why we implemented three-component indexes to solve the full-text search task.

In the future, it will be interesting to investigate other types of queries in more detail and to optimize index creation algorithms for larger values of *MaxDistance*. It will also be important to investigate how the proposed indexing structure can be used by modern ranking algorithms. The author assumes that based on Zipf's law [6], our test text collection is sufficient and acceptable for evaluating search performance. Nevertheless, to investigate ranking algorithms' behavior we plan to use collections, such as TREC GOV and GOV2, which are intended to analyze search quality.

21ing additional indexes for fast full-text searching phrases that contains frequently used words). Control systems and information technologies, **63**(1), 27–33 (2016) (In Russian).
21. Veretennikov, A.B.: Proximity full-text search with a response time guarantee by means of additional indexes. In: Arai K., Kapoor S., Bhatia R. (eds) Intelligent Systems and Applications. IntelliSys 2018. Advances in Intelligent Systems and Computing, vol 868, pp 936-954 (2019). Springer, Cham. doi: 10.1007/978-3-030-01054-6_66.
22. Williams J.W.J.: Algorithm 232 – Heapsort. Communications of the ACM. **7**(6), 347–348, (1964).
23. Jansen, B.J., Spink, A., Saracevic, T.: Real life, real users and real needs: A study and analysis of user queries on the Web. Information Processing and Management, 36(2), 207–227 (2000). doi: 10.1016/S0306-4573(99)00056-4.
# See also

Veretennikov A. B. Using Additional Indexes for Fast Full-Text Searching Phrases that Contain Frequently Used [Control Systems and Information Technologies]. 2013. vol. 52, no. 2. pp. 61-66, In Russian.
**http://veretennikov.org/CLB/Data/clb5en.pdf**

Veretennikov A.B. (2019) Proximity Full-Text Search with a Response Time Guarantee by Means of Additional Indexes. In: Arai K., Kapoor S., Bhatia R. (eds) Intelligent Systems and Applications. IntelliSys 2018. Advances in Intelligent Systems and Computing, vol 868, pp 936-954. Springer, Cham
**https://doi.org/10.1007/978-3-030-01054-6_66**
**http://veretennikov.org/CLB/Data/IntelliSys_2018_ProximityAddInd.pdf**

Veretennikov A.B. Proximity full-text search with a response time guarantee by means of additional indexes with multi-component keys. Selected Papers of the XX International Conference on Data Analytics and Management in Data Intensive Domains (DAMDID/RCDL 2018), Moscow, Russia, October 9-12 2018, 123-130 (2018)
**http://ceur-ws.org/Vol-2277**
**http://veretennikov.org/CLB/Data/DAMDID_2018.pdf**